# On the Feasibility of Attribute-Based Encryption on Internet of Things Devices

Moreno Ambrosin[*], Arman Anzanpour[†], Mauro Conti[*], Tooska Dargahi[‡],
Sanaz Rahimi Moosavi[†], Amir M. Rahmani[†], and Pasi Liljeberg[†]
[*]University of Padua, [†]University of Turku, [‡]CNIT - University of Rome Tor Vergata

**Abstract**—Attribute-Based Encryption (ABE) could be an effective cryptographic tool for the secure management of Internet-of-Things (IoT) devices, but its feasibility in the IoT has been under-investigated thus far. This article explores such feasibility for well-known IoT platforms, namely, Intel Galileo Gen 2, Intel Edison, Raspberry Pi 1 Model B, and Raspberry Pi Zero, and concludes that adopting ABE in the IoT is indeed feasible.

## 1  Introduction

Internet of Things is a growing trend populating the world with billions of interconnected devices. These devices relate to physical "things", ranging from wearable sensors, to smartphones and smart cars [1]. Unfortunately, although IoT has the potential to enable new innovative services and simplify the communication between people and objects, it brings new security and privacy challenges. For example, consider an IP-enabled sensor in a smart healthcare system, which transmits medical data of patients to a remote healthcare server. In this scenario, the conveyed medical data may be routed through an untrusted network, or could be stored in an untrusted cloud service, exposing potentially privacy sensitive data to cyber attacks.

Besides generic IoT security and privacy issues, the concept of *distributed IoT* [1] introduces additional context-specific challenges. Devices not only send their data to the cloud, but also can form an "Intranet of Things", communicating with each other, and with other IoT systems. For example, in a smart healthcare system, devices in a patient's smart house may need to interact directly with hospital's IoT system. However, either collaborating entities may be untrusted, or the transmitted data may need to be revealed only to some selected parties. These challenges call for an urgent need for *efficient authentication* and *fine-grained access control* mechanisms, requiring *advanced cryptographic* methods. Furthermore, an important aspect to consider when it comes to resource-constrained IoT devices is providing flexible key management protocols; which motivated researchers to develop efficient security solutions for IoT systems [2].

**Attribute-Based Encryption and IoT.** In recent years, several security protocols adopted Attribute-Based Encryption (ABE) as a building block in different distributed environments [3], such as IoT [4], cloud services [5], and medical systems [6]. ABE is a public key scheme where both encryption and decryption are based on high-level data access policies. Considering the aforementioned requirements in distributed and heterogeneous IoT scenarios, ABE provides more efficient access control mechanism compared to conventional cryptographic algorithms [3], [6], [7]: (i) allows fine-grained access control based on recipients' attributes; (ii) scales independent from the number of authorized users; (iii) is resilient against collusion attacks; (iv) does not require key sharing or key management algorithms between the participating parties (data owner *does not need* to identify the destination client). Besides its noteworthy advantages, a proper key revocation algorithm is still a challenging issue in ABE (beyond the scope of this paper), and an ongoing research effort [3]. More relevant to our work, ABE suffers from high computational overhead [6], [8]. However, the literature is still missing a proper assessment of ABE efficiency on resource-constrained devices, widely used in the IoT domain.

In order to shine a light on the feasibility of ABE in IoT, we perform a comprehensive analysis of the cost of ABE operations on resource-constrained devices. In particular, along the same line of our previous work [7], which investigated the feasibility of ABE on smartphone devices, in this paper we implement the original Key-Policy Attribute-Based Encryption (KP-ABE) [9] and Ciphertext-Policy Attribute-Based Encryption (CP-ABE) [10] on widely used IoT-enabling devices. Our work focuses on the evaluation of encryption and decryption (hereinafter called *cryptographic operations*) on four boards: Intel Galileo Gen 2, Intel Edison, Raspberry Pi 1 Model B, and Raspberry Pi Zero. Due to space limitation, we only report the results for CP-ABE. However, we noticed that the KP-ABE experiments have a very similar quantitative behavior to CP-ABE results. Supported by our observations from thorough experimental results, we provide evidence of the feasibility of adopting ABE on resource-constrained devices. Moreover, we present a smart healthcare use case application to evaluate feasibility of using ABE in real world IoT scenarios.

## 2    Expressive Encryption with ABE

In KP-ABE, each user's key represents an access policy, e.g., *(Dev_family=Board_XYZ $\wedge$ Dev_role=Role_1) $\vee$ (Release_Date>2013)*, where *Dev_family* and *Dev_role* represent string attributes, *Release_Date* represents a numeric attribute, and $\wedge$ and $\vee$ are the *AND* and *OR* Boolean operators, respectively. Figure 1(a) shows a KP-ABE example, where a data owner encrypts the data specifying a list of attributes. If the data owner assigns the following set of attributes to the ciphertext: *{Dev_family=Board_XYZ, Dev_role=Role_1}* or *{Release_Date=2014}*, the user will be able to decrypt the ciphertext. Because, in these cases the access policy associated to the user's secret key can be satisfied by the attributes assigned to the ciphertext.

Unlike KP-ABE, CP-ABE "enforces" the access policy directly on the data: each user's key is associated with a set of attributes, and a user can decrypt a ciphertext if her attributes satisfy the defined access policy on the data. An example of the CP-ABE is illustrated in Fig. 1(b), where the data owner encrypts the data specifying the access policy: *(Dev_family=Board_XYZ ∧ Dev_role=Role_1) ∨ (Release_Date>2013)* as part of the encryption. A user will be able to decrypt the ciphertext, iif her secret key is associated with a set of attributes that can satisfy the access policy.

Several factors influence the performance of ABE in real world applications, such as desired security level, capacity of the underlying device (i.e., available memory and CPU speed), and the number and type of attributes used in the access policy definition. Attributes number, in particular, plays a fundamental role in ABE performance: encryption in CP-ABE requires computing two exponentiations for each attribute in the resulting access policy. Similarly, KP-ABE encryption requires two exponentiations for each attribute enforced on the ciphertext. Decryption complexity in CP-ABE is upper bounded by $l$ exponentiations, and $2l$ pairing operations [10], while in KP-ABE by only $l$ pairing operations; $l$ is the number of attributes "matching" the access policy (in CP-ABE) or the key policy (in KP-ABE).

For a more complete evaluation of ABE, in this research, we also analyze the impact of using numeric attributes along with string attributes. We believe that, while the use of numeric attributes may be expensive, it provides additional expressiveness in policy definitions, especially in CP-ABE. As an example, there may be situations where access to data should be restricted to only a certain model of devices, released after a certain date (which can be represented as a 64 bit integer).

## 3 Feasibility of ABE on IoT Devices

Despite some researchers' argument about non-acceptable performance of ABE on mobile devices [8], in our previous study [7] we implemented AndrABEn [7], an ABE library for the Android operating system and proved its efficiency. Along the similar line, in this section, we discuss the feasibility of ABE on resource-constrained IoT devices.

Before diving into the results of our experimental analysis, we clarify the concept of "feasibility" that we consider in this paper. As the most important discriminant factor to define feasibility in this domain, we consider *latency*, which has a direct impact on the consumed energy. The results from our study allow to determine, at a high-level, whether the use of ABE is feasible or not in specific applicative scenarios (e.g., video streaming, remote monitoring of healthcare appliances [11]), with respect to their latency requirements. In Section 4 we present a smart healthcare use case example that uses CP-ABE for data encryption. Based on the use case specific latency requirements, we are able to "tune" the adopted security level, and determine the only reasonable number of attributes.

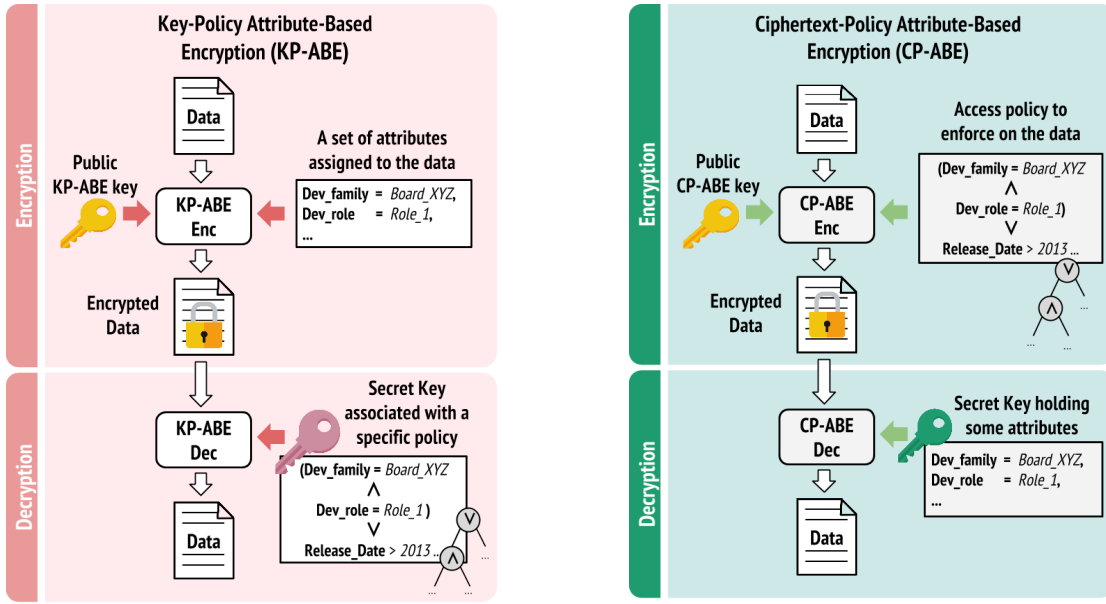

(a) KP-ABE  (b) CP-ABE

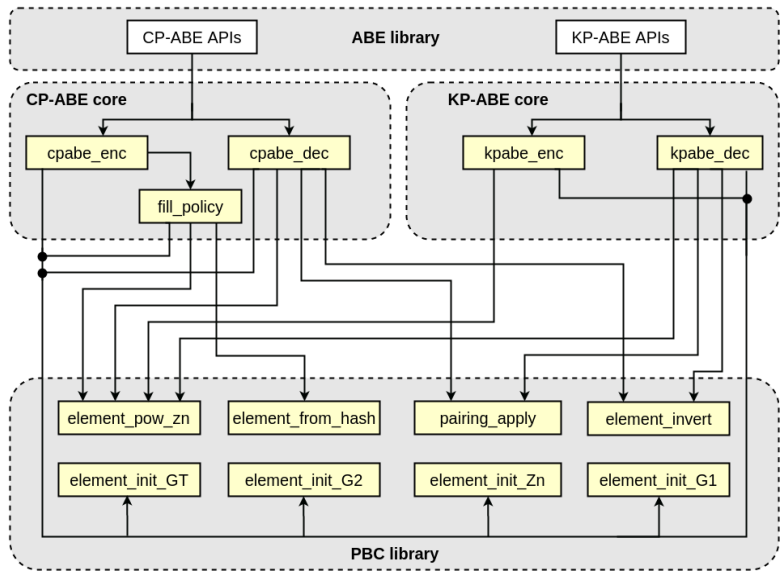

(c) Simplified library structure

**Fig. 1.** High-level overview of KP-ABE and CP-ABE.

## 3.1 Experimental Setup

In our experiments, we adopt the same core C implementation of CP-ABE and KP-ABE used in [7] (code available at: http://spritz.math.unipd.it/projects/andraben), which implement the schemes in [10] and [9], respectively. Figure 1(c) presents a simplified representation of the library, showing its main dependencies with the PBC library of Ben Lynn, at a function-call level. For simplicity, we show only cryptographic operations. Note that, while we are aware of the existence of more recent and improved ABE schemes (such as [3]), we focused on the original schemes for their adoption in previous work [6], [7] to maintain compatibility and comparability, and due to the availability of implementation libraries. We have chosen a set of middle class IoT devices: low cost, with a few megabytes of memory, network-enabled and compatible with a wide range of peripherals, to be used in different industrial or home automation applications [12]. The motivation behind our choice is to explore the performance characteristics of ABE on IoT devices with diverse processing capabilities.

For our evaluation, we used the following setting:
- Intel Edison board: Silvermont Dual Core Intel Atom (500 MHz) + Intel Quark (100 MHz), 632 total DMIPS, 256 MByte memory, Yocto Linux OS, 1335.84 mW baseline power.
- Intel Galileo Gen 2: Intel Quark X1000 (400 MHz), 500 total DMIPS, 1 GByte memory, Yocto Linux OS, 7021.44 mW baseline power.
- Raspberry Pi 1 Model B: ARM1176JZF-S (700 MHz), 875 total DMIPS, 512 MByte memory, Raspbian OS, 2358.4 mW baseline power.
- Raspberry Pi Zero: ARM1176JZF-S (1000 MHz), 1250 total DMIPS, 512 MByte memory, Raspbian OS, 1504 mW baseline power.

We evaluated the performance of cryptographic operations by varying the assured security level, i.e., the number of bits that are used as primitives in cryptographic operations. Longer primitives lead to higher security level. We considered three different security levels (consistent with [7], [8]), equivalent to the security provided by AES symmetric encryption using key lengths of 80, 112, and 128 bits (corresponding to 1024, 2048, and 3072 bits in RSA, respectively). To eliminate the impact of ciphertext size on execution time, we used a symmetric key to encrypt the plaintext, and measured the performance of cryptographic operations of such key. We considered policies having different number of attributes, ranging from 1 to 30. We believe this range represents a reasonable choice in real scenarios, while being consistent with related work [6-8]. Since all the devices run operating systems that support multi-tasking, we report the average execution time for each board collected over several simulations, minimizing the impact of any background tasks on the results.

## 3.2 Evaluation and Discussion

Figures 2 and 3 show execution time, memory usage and energy consumption of CP-ABE on the considered devices, varying number of attributes and security levels (confidence intervals are included in the figures but are not visible since they are too small). It is evident that, as expected, increasing the number of attributes leads to increased execution time and memory usage (and consequently, increased energy consumption). Similarly, a higher security level leads to increased workload on the tested devices. Memory usage footprint is similar for all the boards, ranging between 14 and 15 MByte using a small/medium number of attributes. Security level does not have a significant impact on memory usage, which is rather affected by the number of adopted attributes.

In terms of execution time and energy consumption, Raspberry Pi 1 and Raspberry Pi Zero have similar behavior, and show the best performance, while Intel Galileo shows the worst performance. For example, considering 80 bits security level and 30 attributes, it takes ≈5 sec for encryption, and ≈3.6 sec and ≈2.9 sec for decryption, on Raspberry Pi 1 and Raspberry Pi Zero, respectively. With Intel Galileo, the execution time is ≈15 sec, and ≈13 sec for encryption and decryption, respectively. For a comparison, note that establishing a TLS (version: 1.2; cipher: ECDHE-RSA-AES128-GCM-SHA256; key length: 2048) session with www.google.com:443, on Intel Edison, requires on average 0.206 sec. In the same setting, energy consumption of decryption and encryption on Raspberry Pi 1 and Raspberry Pi Zero are ≈0.5 J, and ≈0.8 J, respectively, while Intel Galileo requires ≈3.7 J and ≈4.3 J, respectively.

Our study provides a clear estimate of how the above two factors contribute to the overall performance, offering a caveat for choosing security level and attributes. In general, performance penalty is higher when increasing the security level, compared to the number of attributes. In order to have stronger security (i.e., moving from 80 to 128 bits), the number of considered attributes needs to be reduced, on average, by 10 times. As an example of the tradeoff between security and number of attributes, CP-ABE encryption with 15 attributes and 112 bits security level shows an average execution time of 9.68 sec, and energy consumption of 1.75 J. Similar performance can be achieved with a security level of 128 bits using policies with less than 5 attributes. A notable insight from our experimentation is this pareto-space of a combinatorial choices of platform, security levels and attributes.

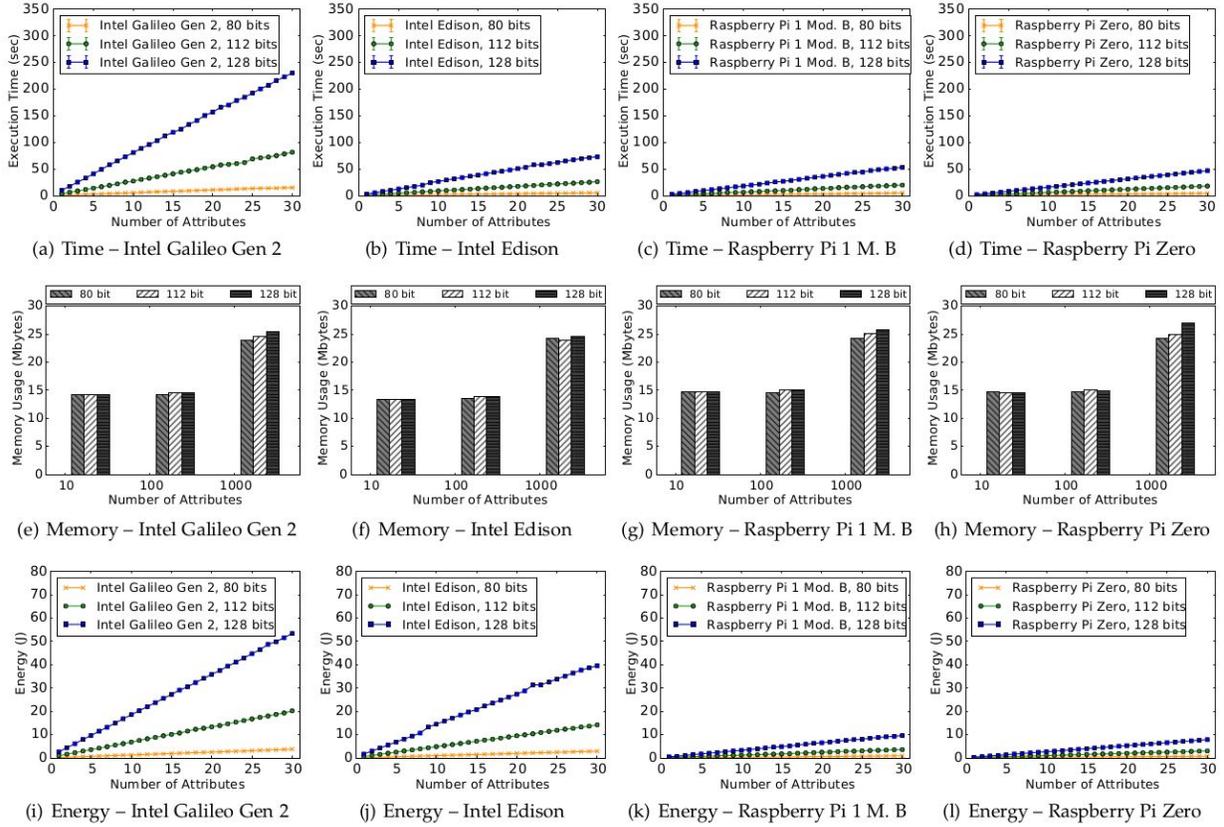

**Fig. 2.** Execution time, memory and energy consumption for CP-ABE encryption.

We further analyzed the overhead of our implementation at a function-call level, i.e., we measured the timing overhead introduced by each function in CP-ABE cryptographic operations, on the Intel Edison board. In general, the encryption routine spends almost 91% of the time executing (multiple times) two functions from the PBC library: element_from_hash, to convert and hash value into a group element, and element_pow_zn, to perform exponentiation in $Z_N$; while decryption depends almost entirely on the pairing_apply function (almost 97% overhead).

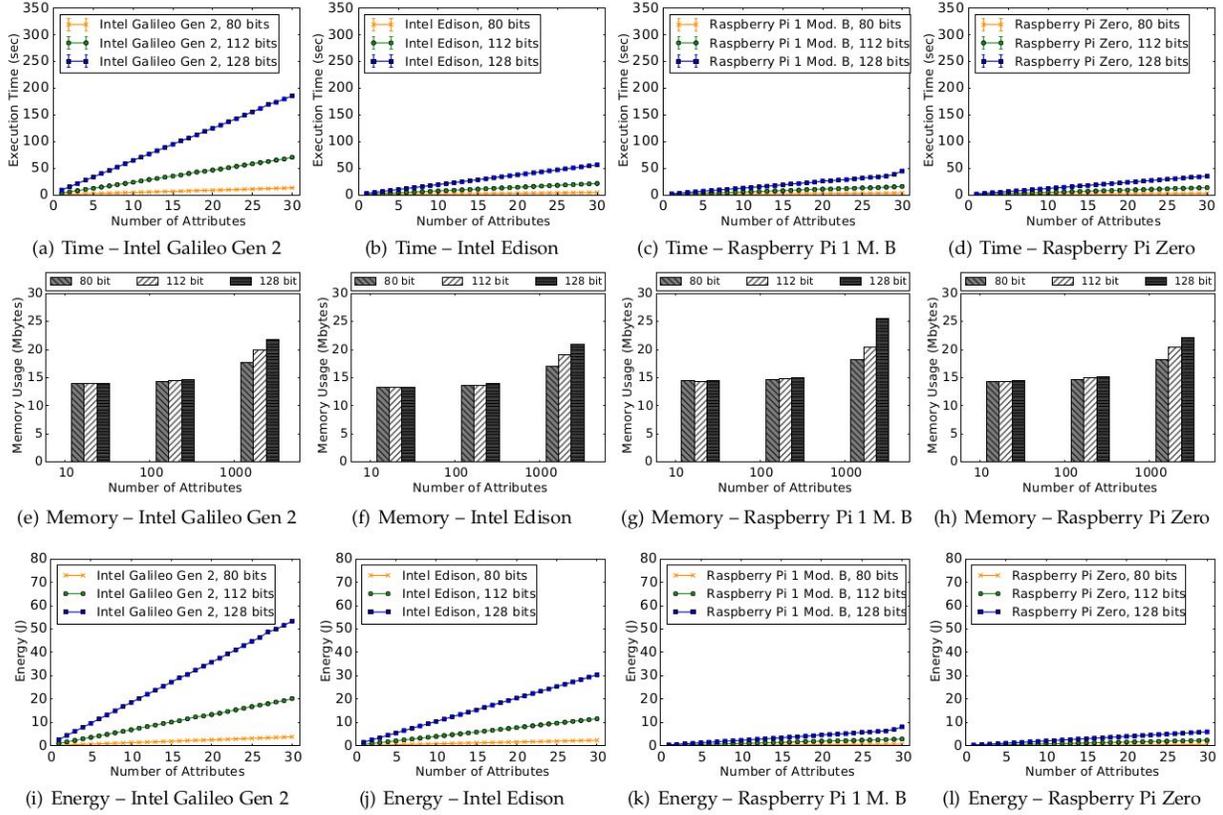

**Fig. 3.** Execution time, memory and energy consumption for CP-ABE decryption.

### 3.3 Numeric Attributes in ABE

According to the original design of CP-ABE [10], access policies are expressed as conjunction of Boolean predicates, e.g., *A* (i.e., *A=true* ), or *A<N*, where *N*∈N, and are represented as trees. Leaf nodes of such trees (e.g., *A, B* and *C* in Figure 4(a)) are attributes, while inner nodes represent *logical threshold gates* of the form *K of N*, meaning that, in order for a set of attributes to satisfy the subtree rooted in such gate, the set must (recursively) satisfy at least *K* of the *N* subtrees of the inner node. A leaf node, i.e., an attribute, is satisfied by a key, if such attribute is associated with the key.

Consider the example in Figure 4(a). The policy: (A∧B)∨C is translated into a tree with three leaves, and two inner threshold gates. The ∧ Boolean operator is translated into a *2 of 2* gate (i.e., both subtrees connected to this gate must be true, in order for this gate to be considered as true), while the ∨ operator as a *1 of 2* gate (i.e., if at least one of the connected nodes to this gate is true, this gate will be considered as true).

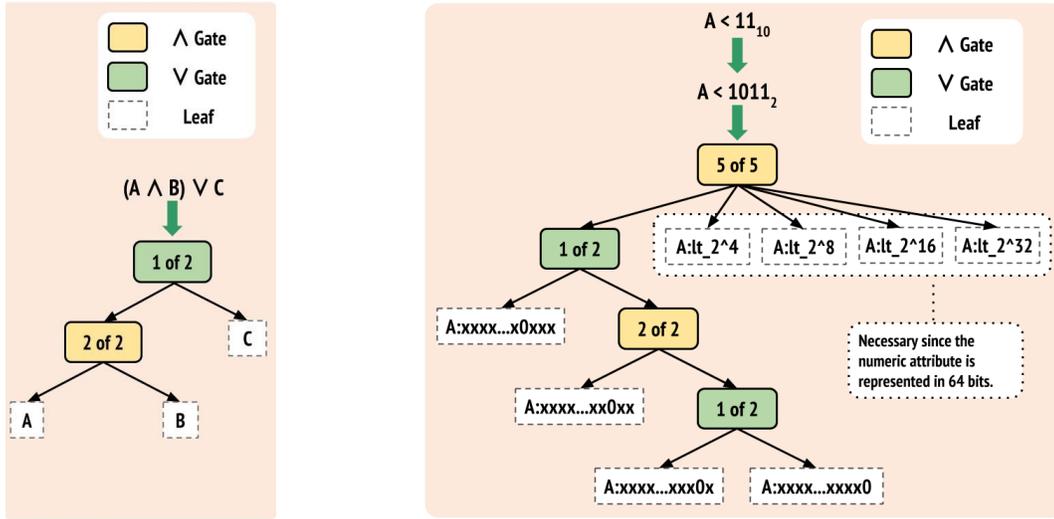

(a) Simple policy            (b) Policy with numeric attributes

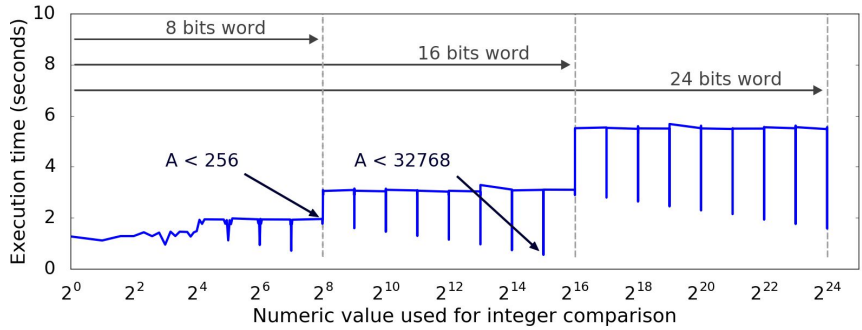

(c) CP-ABE encryption time on a Raspberry Pi 1 Mod B, access policy $A<N$, $N$ ranges from $2^0$ to $2^{24}$.

**Fig. 4.** Access policy translation in CP-ABE.

According to [10], a numeric attribute, such as $A=9$, can be translated into a set of simple attributes indicating the value of each single bit in the attribute's binary representation. For example, using a 64 bit representation for an integer, the attribute $A=9_{10}=1001_2$ is translated into:

$$A\text{:xxxx...1xxx, } A\text{:xxxx...x0xx,}$$
$$A\text{:xxxx...xx0x, } A\text{:xxxx...xxx1,}$$
$$A\text{:eq\_09, } A\text{:gt\_2\textasciicircum02, } A\text{:lt\_2\textasciicircum04, } \ldots$$

This represents the binary translation of *9* (x is a wildcard bit value), plus an attribute for exact matching (A:eq_09), and other attributes, e.g., the ones of the form A:lt_2^N ($A<2^N$), and A:gt_2^N ($A>2^N$), which are "compressed" representations of the remaining bits, required due to the 64 bit representation of a numeric attribute.

Single numeric clauses can be converted into access tree structures of simple attributes. Figure 4(b) shows the translation of *A<11*. As we can see, even simple access control policies involving numeric attributes generate quite complex trees, with a consequent impact on the performance of cryptographic operations. To better understand such impact, we measured the execution time of CP-ABE encryption using simple policies in the form $A<2^X$, where *X* ranges from 1 to 24. Figure 4(c) presents our results, experimented on a Raspberry Pi. We derive two important observations for the practice: (1) encryption time (which depends on the size of the tree) does not grow directly with the size of the considered number, but instead with the "minimum number of Bytes" necessary to represent the number; (2) numbers that are power of *2* generate simpler access trees, with a consequent reduced encryption time. Moreover, for power of 2, the closer the most significant bit at 1 is to the size of the bit word in use (i.e., 8, 16, 24, or 32), the simpler will be the corresponding access tree. For example in Figure 4(c), the access policy *A<256* ($2^8$) generates an access tree with eleven leaves and two AND gates, requiring ≈1.941 sec for encryption; while, encryption with *A<32768* ($2^{15}$) generates a simpler access tree with only three leaves and one AND gate, requiring ≈0.547 sec. Note that, the above considerations on numerical attributes usage can be also extended to the KP-ABE scheme in [9], as it uses a similar access tree construction as in [10].

## 4   Use Case: IoT in Healthcare

To give a "flavor" of the feasibility of using ABE in real world IoT scenarios, we consider a simple, yet realistic use case: smart healthcare. We implemented a prototype wireless healthcare data reader system for remote monitoring, data collection and processing. In our system, measurements from medical sensors are collected, encrypted with CP-ABE, and sent to a data collection server (via Wi-Fi), by an Intel Edison board equipped with an e-Health Sensor Shield V2.0. The whole process is carried out by two services running on the board: the first reads the data from sensors, and writes it into files (one per data type); the second encrypts the files with CP-ABE, and sends them to the server, which may represent an untrusted gateway, cloud service, or another IoT device. Figure 5(a) summarizes our application parameters. The specific system sampling rate requirements give us clear latency constraints based on which one should choose the acceptable range for the number of attributes and security level.

In general, reading and sending rate should be roughly the same, in order to guarantee the expected Quality of Service. Furthermore, as most of the traffic in our scenario consists of ECG data, approximately 1500 Bytes/sec (500 reads of 3 Bytes every second), in what follows we focus on ECG data. Given the ≈80 msec for data transmission (per UDP packet), and the average 45 msec to encrypt the measurements file with AES, the most expensive operations are related to CP-ABE. In order to find a reasonable balance between assured security level, and expressiveness (in terms of attributes number), we carried out tests using up to 10 attributes and 80 bit security level, measuring the overall latency. Referring to Figure 5(b), latency remains smaller, or close to one second (our upper bound for latency) with a maximum of 5 attributes. We can conclude that, CP-ABE can be used in such scenario supporting up to 5 attributes with 80 bits security. Note that encryption time is a bit longer compared to the results in Section 3.2, because: (1) time includes AES encryption and per-file key generation; and (2) the background reading service is always busy recording data.

## 5   Conclusion

We have shown the feasibility of adopting ABE in representative IoT systems. Our results can be a reference for researchers and designers of novel ABE-based security solutions. We believe future research should focus on improving ABE efficiency, via both a careful selection of attributes and software and hardware optimizations for the cryptographic library. Our analysis shows that the utilized library can be significantly optimized via proper memory management, customized data structure deployment, and simplification of cryptographic arithmetic operations considering input attributes. Moreover, considering the fact that the complexity of CP-ABE and KP-ABE depends on the number of exponentiations and pairing operations performed by each of their algorithms, future work could address the migration of complex arithmetic operations, such as exponentiation, to hardware accelerators (for example, custom logic on field-programmable gate arrays) in order to enhance energy efficiency and total execution time

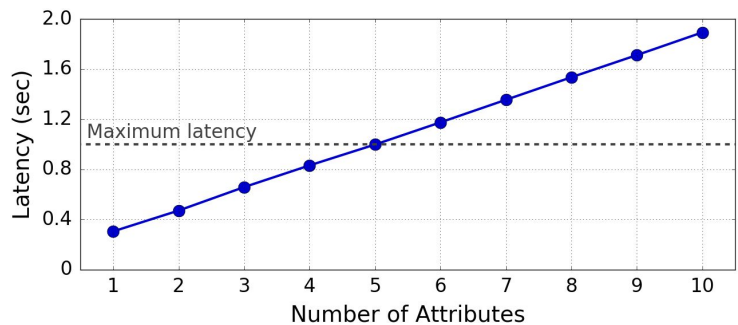

(a) Sensor properties and application parameters

(b) Latency on Intel Edison

**Fig. 5.** Healthcare use case parameters and latency evaluation on an Intel Edison board, considering 80 bits security level.

## Acknowledgement

This research was partially supported by the EU Marie Curie Fellowship PCIG11-GA- 2012-321980 and EU projects ReCRED (ref. 653417), EU TagItSmart! (H2020-ICT30-2015-688061), and EU-India REACH (ICI+/2014/342-896).